\documentclass[10pt, conference]{IEEEtran}
\usepackage{cite}
\usepackage{amsmath}
\usepackage{graphicx}
\usepackage{amssymb}
\usepackage{amsthm}
\usepackage{hyperref}
\usepackage{cleveref}
\usepackage{algorithm}
\usepackage{pgfplots}
\usepackage[noend]{algpseudocode}
\algrenewcommand{\algorithmiccomment}[1]{\hskip3em// \textit{#1}}

\newtheorem{example}{Example}

\providecommand{\keywords}[1]
{
  \textbf{\textit{Keywords---}} #1
}

\begin{document}

\title{SK-Tree: a systematic malware detection algorithm  \\ on streaming trees via the signature kernel}

\author{\IEEEauthorblockN{Thomas Cochrane, Peter Foster, \\ and Varun Chhabra}
\IEEEauthorblockA{The Alan Turing Institute\\
\href{mailto:{thomasc,pfoster,vchhabra}@turing.ac.uk}{\{thomasc,pfoster,vchhabra\}@turing.ac.uk}}
\and
\IEEEauthorblockN{Maud Lemercier}
\IEEEauthorblockA{University of Warwick\\
The Alan Turing Institute\\
\href{mailto:maud.lemercier@warwick.ac.uk}{maud.lemercier@warwick.ac.uk}}
\and
\IEEEauthorblockN{Terry Lyons and Cristopher Salvi}
\IEEEauthorblockA{University of Oxford\\
The Alan Turing Institute\\
\href{mailto:{tlyons,salvi}@maths.ox.ac.uk}{\{tlyons,salvi\}@maths.ox.ac.uk}}}


%


\maketitle

\begin{abstract}
The development of machine learning algorithms in the cyber security domain has been impeded by the complex, hierarchical, sequential and multimodal nature of the data involved. In this paper we introduce the notion of a \emph{streaming tree} as a generic data structure encompassing a large portion of real-world cyber security data. Starting from host-based event logs we represent computer processes as streaming trees that evolve in continuous time. Leveraging the properties of the \emph{signature kernel}, a machine learning tool that recently emerged as a leading technology for learning with complex sequences of data, we develop the \emph{SK-Tree algorithm}. SK-Tree is a supervised learning method for systematic malware detection on streaming trees that is robust to irregular sampling and high dimensionality of the underlying streams. We demonstrate the effectiveness of SK-Tree to  detect malicious events on a portion of the publicly available DARPA OpTC dataset, achieving an AUROC score of $98\%$.

\end{abstract}

\vspace{0.5cm}


%
\IEEEpeerreviewmaketitle

\section{Introduction}

The design and deployment of sophisticated cyber-attacks such as advanced persistent threats \cite{chen2014study} has grown dramatically over the last few years. This has been facilitated by the appearance of new varieties of malware, and new adversary tactics and techniques, which are designed to evade existing defensive products used by enterprises such as anti-virus, firewalls and intrusion detection systems. Modern cyber security systems are generally rule-based and rely on a team of security analysts monitoring network activities and manually investigating suspected malicious activities, to determine the scope of the potential threats. This investigation phase is particularly labour-intensive. In order to defend against the influx of new malware variants and increasingly sophisticated attacks, it is imperative to develop systematic mechanisms to detect them. 

One of the main challenges in creating effective and systematic malware detection systems is the complex nature of the data. The relevant datasets consist of \emph{multimodal streams} of information, i.e. sequences of data generated by a large set of heterogeneous sources (servers, routers, workstations etc.), and representing various types of activity such as logons, file accesses, and network connections. Such data streams are often recorded at irregular intervals, span different time periods and exhibit missing observations, all aspects that make the design of systematic methods even more complicated. 

\begin{figure*}
    \centering
    \includegraphics[scale=0.4]{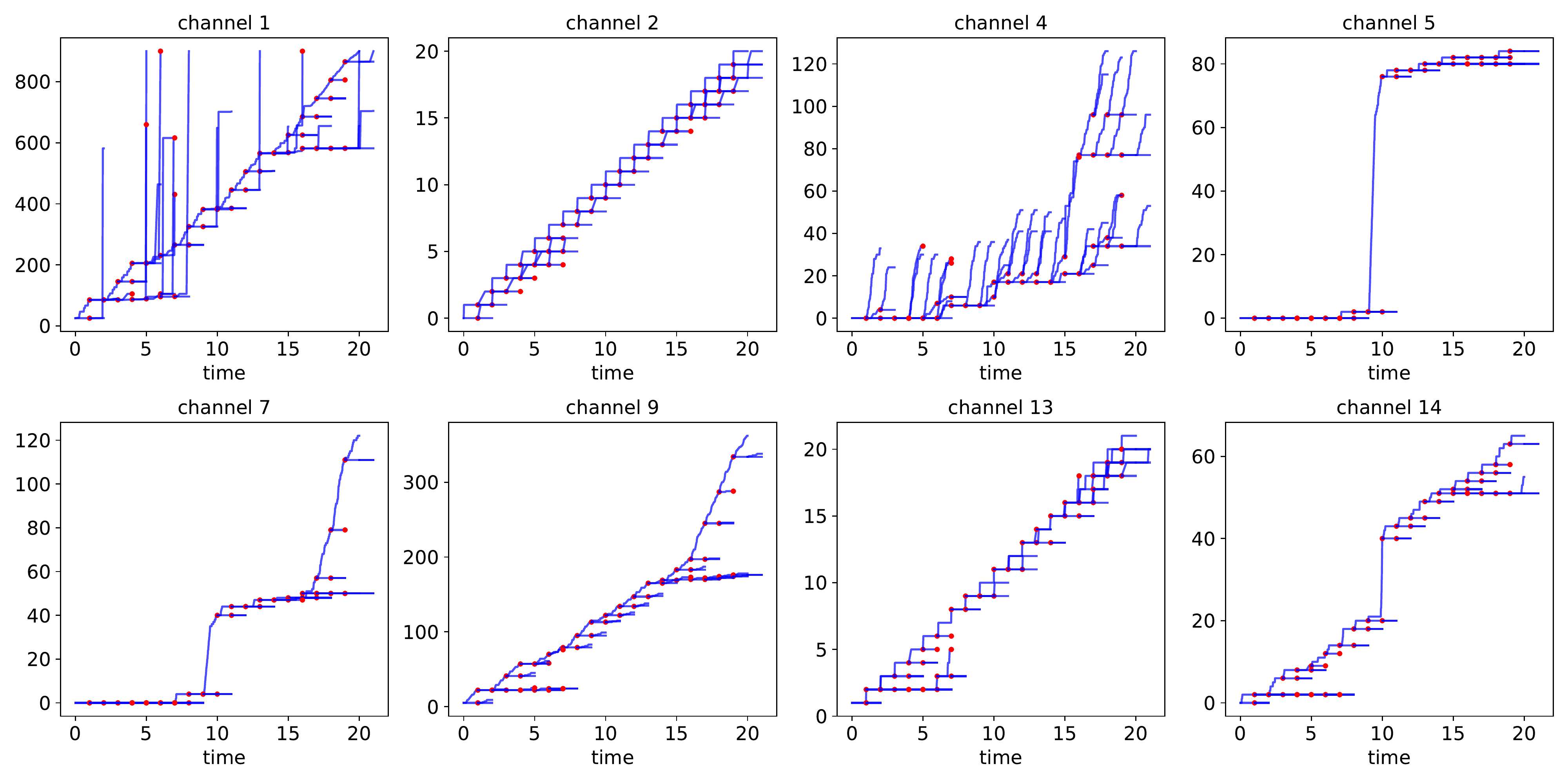}
    \caption{Visual representation of selected channels of one single streaming tree.  Each plot represents the evolution in time of the value of a given channel of the streaming tree, on its various branches. A red dot indicates a point where the currently-tracked process sets off a child process, causing the tree to branch.}
    \label{fig:trees}
\end{figure*}

A common characteristic of this data that cannot be ignored in the analysis is their \emph{hierarchical structure}. For example, a given process may set off child processes, which themselves may spawn more children, producing a \emph{process tree}. In this paper we account for how activity occurs within this tree-based structure over time.  Specifically we represent the behaviour of computer processes, as observed in host-based event logs, as \emph{streaming trees} evolving in continuous time. A representation of some selected channels of a single streaming tree is depicted in Fig. \ref{fig:trees}. The notion of a streaming tree we introduce is a fairly  generic data structure encompassing a large portion of real-world sequential data encountered in the cyber security domain. The structure of a streaming tree is significantly more complex than that of a multivariate time series, as it accounts for the hierarchy of the data; to our knowledge no systematic method has been proposed in the literature (summarised in Sec. \ref{sec:rel_work}) to deal with such a data structure. 

Our main contribution (Sec. \ref{sec:method}) is a systematic malware detection algorithm on streaming trees that we call \emph{SK-Tree}. The core component of SK-Tree is the \emph{signature kernel}, a machine learning tool introduced in \cite{kiraly2019kernels} that can process complex sequences of data. The signature kernel is based on the \emph{path-signature} \cite{chevyrev2016primer,lyons2007differential} which is a well-known transform from stochastic analysis that recently emerged as a leading technology for learning with time series data. SK-Tree is to our knowledge the first systematic malware classification algorithm on streaming trees that is robust to irregular sampling and missing data. We test SK-Tree (Sec. \ref{sec:experiments}) on a portion of the DARPA’s Operationally Transparent Cyber dataset \cite{optc}, which is available openly and one of the largest datasets released to date, achieving an AUROC score of $98\%$. We make our python implementation of SK-Tree publicly available at \url{https://github.com/crispitagorico/SK-Tree}.




\section{Related work}\label{sec:rel_work}
Many machine learning methods developed in the cyber security literature have focused on the processing of network data, with only few techniques having been designed to consume host event data as we do in this work. Among publicly-available datasets, two released by Los Alamos National Laboratory (LANL) are the most widely used \cite{lanl2015cyberdata1,lanl2018unified}.  As far as we are aware, our work is the first to make use of the DARPA OpTC dataset \cite{optc}. Compared to earlier datasets, events in OpTC give a much more detailed summary of host activity, and contain more varied red-team behaviour. The heterogeneity of the data is one of the major bottlenecks.






On the one hand, some of the existing techniques aim to detect malicious activity from streams of events without taking into account any hierarchical structure. There has been a proliferation of work in anomaly detection for the purposes of finding malicious activity \cite{walker2020malware, whitehouse2016activity, brown2018recurrent, riddle2018adaptive, eren2020multi, tuor2017deep,passino2020classification}. In particular \cite{brown2018recurrent} took a host-based approach and considered the authentication events in the 2015 LANL dataset using a RNN language model.  Auto-encoders were used in \cite{holt2019deep} to recognise lateral movement through a network manually engineer features based on conditional probabilities defined by event counts. 

On the other hand, another line of thoughts tries to detect malware or suspicious behaviour using the graph-like structure of the data, but without taking into account the sequential nature of the events. Utilised features include connections between computers on a network or relations of processes to those that they spawn. Tree and graph kernels were first introduced for malware analysis in \cite{wagner2009malwareanalysis}, comparing the similarity of subtrees and subgraphs respectively, and achieving promising results on data obtained from a honeypot. Process trees were further studied in \cite{wijands2015processtree}, where structures from process trees were compared against reference trees known to be non-malicious. A similar technique was used by \cite{luh2020advancedthreat}, comparing star structures with templates extracted from uninfected trees. 

However, none of the aforementioned approaches are designed to process complex data structures that are hierarchical and sequential simultaneously. The main objective of this paper is to provide a general procedure flexible enough to handle such complex data structures, formalised in the next section. 


\section{Method}\label{sec:method}

\subsection{Streaming trees}\label{sec:trees}

A \emph{multivariate time series $\mathbf{x}$ of dimension $d \in \mathbb{N}$} is a collection of points $x_i \in \mathbb{R}^{d-1}$ with corresponding time-stamps $t_i \in \mathbb{R}$ such that $t_0 < ... < t_n$ and defined as follows
\begin{equation}\label{eqn:time-series}
    \mathbf{x} = ((t_0,x_0), (t_1, x_1), ..., (t_n, x_n))
\end{equation}
A \emph{tree} $\tau$ is defined recursively as a tuple of the form $\tau = (\mathbf{x}_\tau,F_\tau)$, where $\mathbf{x}_\tau$ is a multivariate time series and $F_\tau=(\tau_1, ..., \tau_k)$ is a (possibly empty) collection of trees. Concatenating to the time series $\mathbf{x}_\tau$ of a tree $\tau$ the values of the time series $\mathbf{x}_{\tau_1}, ..., \mathbf{x}_{\tau_k}$ of its children subtrees $\tau_1, ..., \tau_k$, and repeating this operation recursively for each children subtree $\tau_i$, we end up with an equivalent representation of the tree $\tau$ as a collection of branches $\tau= (\mathbf{x}_\tau^1, ...,\mathbf{x}_\tau^n)$, where each branch $\mathbf{x}_\tau^i$ is a multivariate time series starting at $t_0$. For the sequel, it is important to keep both representations in mind.

\begin{example}\label{ex:tree}
\normalfont Consider the tree $\tau = (\mathbf{x}_\tau, ((\mathbf{x}_{\tau_1}, \varnothing), (\mathbf{x}_{\tau_2}, \varnothing)))$, where $\varnothing$ denotes the empty list and where $\mathbf{x}_\tau, \mathbf{x}_{\tau_1}, \mathbf{x}_{\tau_2}$ are the following multivariate time series of dimension $d$
\begin{align*}
    &\mathbf{x}_\tau = ((t_0,x_0), (t_1, x_1), ..., (t_n, x_n)) \\
    &\mathbf{x}_{\tau_1} = ((t_{n+1},y_0), (t_{n+2}, y_1), ..., (t_{n+i}, y_{i-1})) \\
    &\mathbf{x}_{\tau_2} = ((t_{n+1},z_0), (t_{n+2}, z_1), ..., (t_{n+j}, z_{j-1}))
\end{align*}
It is easy to see that $\tau$ has two branches. Indeed, consider a first multivariate time series of dimension $d$ 
\begin{align*}
    \mathbf{x}_\tau^1 = ((t_0,x_0), (t_1, x_1), ..., (t_{n+i}, x_{n+i}))
\end{align*}
where 
\begin{equation*}
    x_k = \begin{cases}
            x_k, &         \text{if } k \leq n ,\\
            y_{k-n-1}, & \text{if }  n+1\leq k \leq n+i
    \end{cases}
\end{equation*}
and a second multivariate time series of dimension $d$ 
\begin{align*}
    &\mathbf{x}_\tau^2= ((t_0,x_0), (t_1, x_1), ..., (t_{n+j}, x_{n+j})) 
\end{align*}
where 
\begin{equation*}
    x_k = \begin{cases}
            x_k, &         \text{if } k \leq n ,\\
            z_{k-n-1}, & \text{if }  n+1\leq k \leq n+j
    \end{cases}
\end{equation*}
Then, $\tau$ has the equivalent representation $\tau=(\mathbf{x}^1_{\tau}, \mathbf{x}^2_{\tau})$ in terms of its two branches $\mathbf{x}^1_{\tau}, \mathbf{x}^2_{\tau}$.
\end{example}


Given a multivariate time series $\mathbf{x}$ of the form of \cref{eqn:time-series}, define the \emph{path} $X : [t_0,t_n] \to \mathbb{R}^d$ as the the continous piecewise linear interpolation of the data such that $X_{t_i} = (t_i, x_i)$. A \emph{streaming tree} $\mathcal{T}$ is the data structure obtained by replacing all the time series appearing in the definition of a tree $\tau$ (and in all its children subtrees) by their continuous piecewise linear interpolation. Analogously to the equivalent representation of a tree in terms of its branches that we discussed above, a streaming tree $\mathcal{T}$ can also be represented as a collection of branches $\mathcal{T} = (X^1_\mathcal{T}, ..., X^n_\mathcal{T})$ where each branch $X^i_\mathcal{T}$ is the continuous piecewise linear interpolation of the $i^{th}$ branch $\mathbf{x}_\tau^i$ of the tree $\tau$. The resulting branches $X^1_\mathcal{T}, ..., X^n_\mathcal{T}$ form a collection of paths with common history.

\begin{example}
\normalfont
Consider again the same tree as in Example \ref{ex:tree}, i.e. $\tau = (\mathbf{x}_\tau, ((\mathbf{x}_{\tau_1}, \varnothing), (\mathbf{x}_{\tau_2}, \varnothing)))$. We saw that $\tau$ can be represented in terms of its two branches as $\tau=(\mathbf{x}_{\tau}^1, \mathbf{x}_{\tau}^2)$. Let $X^1_\mathcal{T}:[t_0,t_{n+i}] \to \mathbb{R}^d$ and $X^2_\mathcal{T}:[t_0,t_{n+j}] \to \mathbb{R}^d$ be the continuous piecewise linear interpolation of $\mathbf{x}_{\tau}^1, \mathbf{x}_{\tau}^2$ respectively. Then the streaming tree $\mathcal{T}$ can be represented in terms of its two branches as $\mathcal{T}=(X^1_\mathcal{T},X^2_\mathcal{T})$.
\end{example}

\subsection{The signature}

Here we describe a well-known path-transform called the \emph{signature} that allows us to extract meaningful features from a multivariate time series in a systematic way. The signature has been deployed as a machine learning tool in many data science applications dealing with sequential data \cite{arribas2018signature, kalsi2020optimal, moore2019using, li2019skeleton}.

For any coordinate $\alpha \in \{1,...,d\}$ and any continuous piecewise linear path $X:[0,T] \to \mathbb{R}^d$ we denote its $\alpha^{th}$ \emph{channel} by $X^{(\alpha)}$ so that at any time $t \in [0,T]$
\begin{equation}
    X(t) = (X^{(1)}(t), ..., X^{(d)}(t)). 
\end{equation}
We denote by $\mathcal{X}$ the set of all continuous piecewise linear paths with values in $\mathbb{R}^d$. The signature $S:\mathcal{X} \to H$ is a \emph{feature map} defined for any path $X \in \mathcal{X}$ as the following infinite collection of features \cite{chevyrev2016primer}
\begin{align}
    S(X) = \Big(1, &\left\{S(X)^{(\alpha_1)}\right\}_{\alpha_1=1}^d, \nonumber \\
    &\left\{S(X)^{(\alpha_1,\alpha_2)}\right\}_{\alpha_1,\alpha_2=1}^d, \nonumber \\
    &\left\{S(X)^{(\alpha_1,\alpha_2,\alpha_3)}\right\}_{\alpha_1,\alpha_2,\alpha_3=1}^d,... \Big)
\end{align}
where every term is a scalar defined as the iterated integral 
{\small
\begin{equation}\label{eqn:sig}
S(X)^{(\alpha_1,...,\alpha_j)} =  \underset{0<s_1<...<s_j<T}{\int ... \int} d X^{(\alpha_1)}(s_1) ... dX^{(\alpha_j)}(s_j)
\end{equation}
}%
The \emph{feature space} $H$ associated to the signature is a Hilbert space defined as the direct sum of tensor powers of $\mathbb{R}^d$
\begin{equation}
    H = \bigoplus_{k=0}^\infty (\mathbb{R}^{d})^{\otimes k} = \mathbb{R} \oplus \mathbb{R}^{d}\oplus (\mathbb{R}^{d})^{\otimes 2} \oplus ... 
\end{equation}
where $\otimes$ denotes the outer product \cite{lyons2007differential}.\medskip

\subsection{The expected signature}

Here we describe another transform for sequential data called the \emph{expected signature} that generalises the signature in the sense that it allows to extract useful features from an ensemble of time series. The sequence of moments $(\mathbb{E}[Z^{\otimes m}])_{m\geq 0}$ of any finite dimensional random variable $Z$ is classically known to characterize its law $\mu_Z = \mathbb{P} \circ Z^{-1}$. It turns out that in the infinite dimensional case of path-valued random variables an analogous result holds \cite{chevyrev2018signature}; it says that one can fully characterise such path-valued random variables by replacing moments by the expected signature, that we define next. 

Assume that $\mathcal{X}$ is compact and let $\mu$ be a probability measure supported on $\mathcal{X}$. The \emph{expected signature} of $\mu$ is defined as the following infinite collection of statistics  
\begin{align*}
    \mathbb{E}_S(\mu) = \Big(1, &\left\{\mathbb{E}_S(\mu)^{(\alpha_1)}\right\}_{\alpha_1=1}^d,\\
    &\left\{\mathbb{E}_S(\mu)^{(\alpha_1,\alpha_2)}\right\}_{\alpha_1,\alpha_2=1}^d, \\
    &\left\{\mathbb{E}_S(\mu)^{(\alpha_1,\alpha_2,\alpha_3)}\right\}_{\alpha_1,\alpha_2,\alpha_3=1}^d,... \Big)
\end{align*}
where each term is a scalar defined as the following integral
\begin{equation}
    \mathbb{E}_S(\mu)^{(\alpha_1, ..., \alpha_j)} = \int_{X \in \mathcal{X}}S(X)^{(\alpha_1, ..., \alpha_k)}\mu(dX).\medskip
\end{equation}

As explained in \cref{sec:trees}, a streaming tree $\mathcal{T}$ can be represented in terms of its branches as a collection of continuous piecewise linear paths $\mathcal{T}=(X^1_\mathcal{T}, ..., X^n_\mathcal{T})$. Denoting by $\delta_{X^i_\mathcal{T}}$ the \emph{Dirac measure} indexed on the path $X^i_\mathcal{T}$ we can represent the streaming tree $\mathcal{T}$ as the following \emph{empirical measure}
\begin{equation}
    \mu_{\mathcal{T}} = \frac{1}{n}\sum_{i=1}^n \delta_{X^i_\mathcal{T}}
\end{equation}
We refer the reader to \cite{lemercier2020distribution} for further details on the probabilistic setup. Therefore, any streaming tree $\mathcal{T}$ can be faithfully represented by means of its expected signature $\mathbb{E}_S(\mu_{\mathcal{T}})$. Given the special recursive structure of a streaming tree as collection of paths with common history, the expected signature $\mathbb{E}_S(\mu_{\mathcal{T}})$ can be computed via a convenient recursive formula: consider a streaming tree $\mathcal{T} = (X_\mathcal{T}, F_\mathcal{T})$, where $F_\mathcal{T}=(\mathcal{T}_1, ..., \mathcal{T}_n)$ is a (possibly empty) list of streaming trees. Then, the expected signature $\mathbb{E}_S(\mu_{\mathcal{T}})$
satisfies the following recursive formula
\begin{equation}
\mathbb{E}_S(\mu_{\mathcal{T}}) = 
    \begin{cases}
            S(X_\mathcal{T}), &         \text{if } F_\mathcal{T} = \varnothing ,\\
            \frac{1}{n} \sum_{i=1}^n S(X_\mathcal{T}) \otimes \mathbb{E}_S(\mu_{\mathcal{T}_i}) , & \text{otherwise } 
    \end{cases}
\end{equation}
where $S(X_\mathcal{T})$ is the signature of the path $X_\mathcal{T}$, $\varnothing$ is the empty list, $\otimes$ is the outer product and $\mathbb{E}_S(\mu_{\mathcal{T}_i})$ is the expected signature of the streaming tree $\mathcal{T}_i$.

\subsection{A measure of similarity between streaming trees}

Consider two streaming trees $\mathcal{T}_1$ and $\mathcal{T}_2$ with corresponding probability measures $\mu_{\mathcal{T}_1}$ and $\mu_{\mathcal{T}_2}$. An appropriate measure of similarity between the trees $\mathcal{T}_1$ and $\mathcal{T}_2$ can be obtained by considering a distance between the probability measures $\mu_{\mathcal{T}_1}$ and $\mu_{\mathcal{T}_2}$. The \emph{maximum mean discrepancy} (MMD) distance between $\mu_{\mathcal{T}_1}$ and $\mu_{\mathcal{T}_2}$ is defined as
\begin{equation}\label{eqn:MMD}
    d_{\text{MMD}}(\mu_{\mathcal{T}_1}, \mu_{\mathcal{T}_2}) = \sup_{f \in \mathcal{G}} \left| \mathbb{E}_{\mu_{\mathcal{T}_1}}[f(X_{\mathcal{T}_1})] - \mathbb{E}_{\mu_{\mathcal{T}_2}}[f(X_{\mathcal{T}_2})] \right|
\end{equation}
where $X_{\mathcal{T}_i} \sim \mu_{\mathcal{T}_i}$ is a sample path from the probability measure $\mu_{\mathcal{T}_i}$ (for $i=1,2$), where $\mathcal{G}$ is a space of real valued functions on the path space $\mathcal{X}$ defined as the unit ball of the \emph{reproducing kernel Hilbert space} (RKHS) $\mathcal{H}_k$ associated to an appropriate \emph{kernel on paths} $k: \mathcal{X} \times \mathcal{X} \to \mathbb{R}$, in other words
\begin{equation}
    \mathcal{G}=\{f: \mathcal{X} \to \mathbb{R} : ||f||_{\mathcal{H}_k} \leq 1\}
\end{equation}
For a detailed account of RKHSs and MMD distance we refer the interested reader to \cite{gretton2012kernel}. Thanks to the main result in \cite{gretton2012kernel}, the MMD distance of equation (\ref{eqn:MMD}) can be expressed as the sum of the following three terms 
\begin{align}
    d_{\text{MMD}}(\mu_{\mathcal{T}_1}, \mu_{\mathcal{T}_2})^2  & =  \mathbb{E}_{(\mu_{\mathcal{T}_1}, \mu_{\mathcal{T}_1})}[k(X_{\mathcal{T}_1},\widetilde X_{\mathcal{T}_1})]  \nonumber \\
    & + \mathbb{E}_{(\mu_{\mathcal{T}_2}, \mu_{\mathcal{T}_2})}[k(X_{\mathcal{T}_2},\widetilde X_{\mathcal{T}_2})]  \nonumber \\
    & - 2 \mathbb{E}_{(\mu_{\mathcal{T}_1}, \mu_{\mathcal{T}_2})}[k(X_{\mathcal{T}_1},X_{\mathcal{T}_2})] 
\end{align}
where each term is expressed in terms of sample paths from the underlying measures $\mu_{\mathcal{T}_1}, \mu_{\mathcal{T}_2}$ and kernel evaluations on those samples.
If the streaming tree $\mathcal{T}_1 = (X^1_{\mathcal{T}_1}, ..., X^m_{\mathcal{T}_1})$ has $m$ branches and the streaming tree $\mathcal{T}_2 = (X^1_{\mathcal{T}_2}, ..., X^n_{\mathcal{T}_2})$ has $n$ branches, the MMD distance can be computed explicitly as
\begin{align}\label{eqn:estimator}
    d_{\text{MMD}}(\mu_{\mathcal{T}_1}, \mu_{\mathcal{T}_2})^2 & = \frac{1}{m(m-1)}\sum_{i=1}^m\sum_{j\neq i}^m k(X^i_{\mathcal{T}_1}, X^j_{\mathcal{T}_1}) \nonumber \\
    & + \frac{1}{n(n-1)}\sum_{i=1}^n\sum_{j\neq i}^n k(X_{\mathcal{T}_2}^i, X_{\mathcal{T}_2}^j)  \nonumber \\
    & - \frac{2}{mn}\sum_{i=1}^m\sum_{j=1}^n k(X^i_{\mathcal{T}_1}, X^j_{\mathcal{T}_2}) 
\end{align}
Therefore, in order to quantify the similarity between two streaming trees $\mathcal{T}_1, \mathcal{T}_2$ using (\ref{eqn:estimator}) it is paramount to specify an appropriate kernel $k$ acting on paths from $\mathcal{X}$.

\subsection{The signature kernel}\label{ssec:signaturekernel}

The \emph{signature kernel} $k:\mathcal{X}\times\mathcal{X} \to \mathbb{R}$ is a reproducing kernel associated to the signature feature map $S$ and defined for any pair of paths $X : [0,T] \to \mathbb{R}^d$ and $Y : [0,T] \to \mathbb{R}^d$ as the following inner product \cite{chevyrev2018signature}
\begin{align}\label{eq:kernel}
    k(X,Y) = \left\langle S(X), S(Y) \right\rangle_H 
\end{align} 
In the recent article \cite{cass2020computing} the authors show that the signature kernel can be computed explicitly; they provide a \emph{kernel trick} for the signature kernel by proving the following relation
\begin{equation}\label{eq:pde_kernel}
    k(X,Y)=U(T,T)
\end{equation}
where the function $U : [0,T] \times [0,T] \to \mathbb{R}$ is the solution of the following \emph{partial differential equation} (PDE)
\begin{align}\label{eq:sig_PDE}
    \frac{\partial^2 U}{\partial s \partial t} = \frac{\partial^2 \kappa(X(s), Y(t))}{\partial s \partial t} U
\end{align}
with boundary conditions $U(0,\cdot)=1$ and $U(\cdot,0)=1$, where $\kappa : \mathbb{R}^d \times \mathbb{R}^d \to \mathbb{R}$ is any base kernel on $\mathbb{R}^d$. In this paper we will be using the \emph{Gaussian RBF kernel} as base kernel $\kappa$. Therefore, the MMD distance between two streaming trees $\mathcal{T}_1, \mathcal{T}_2$ can be computed explicitly via equation (\ref{eqn:estimator}). We will make use of this distance in our malware detection algorithm SK-Tree that we introduce next.

\subsection{SK-Tree: a kernel based malware detection algorithm}

In \cite{lemercier2020distribution} the authors construct a universal kernel indexed on probability measures on paths defined as a combination of the signature kernel, MMD distance of equation (\ref{eqn:MMD}), and a Gaussian kernel. Given the representation of a streaming tree as a probability measure on paths, we can leverage their construction and propose the following kernel for streaming trees: let $\mathcal{T}_1 = (X^1_{\mathcal{T}_1}, ..., X^m_{\mathcal{T}_1})$ and $\mathcal{T}_2=(X^1_{\mathcal{T}_1}, ..., X^n_{\mathcal{T}_1})$ be two streaming trees with associated probability measures $\mu_{\mathcal{T}_1}, \mu_{\mathcal{T}_2}$ respectively. We define the kernel $k_{\sigma}$ on streaming trees as
\begin{equation}
    k_\sigma(\mu_{\mathcal{T}_1},\mu_{\mathcal{T}_2})=\exp \left( -\sigma^2 d_{\text{MMD}}(\mu_{\mathcal{T}_1},\mu_{\mathcal{T}_2})^2\right)
\end{equation}
where $\sigma>0$ is a scalar hyperparameter. By equation (\ref{eqn:estimator}), the kernel $k_\sigma$ can be explicitly computed by evaluating the signature kernel $k$ at the branches of $\mathcal{T}_1, \mathcal{T}_2$ as follows
\begin{align}
    k_\sigma(\mu_{\mathcal{T}_1},\mu_{\mathcal{T}_2})  =\exp \Big( & - \frac{\sigma^2}{m(m-1)}\sum_{i=1}^m\sum_{j\neq i}^m k(X^i_{\mathcal{T}_1}, X^j_{\mathcal{T}_1}) \nonumber \\
    & - \frac{\sigma^2}{n(n-1)}\sum_{i=1}^n\sum_{j\neq i}^n k(X^i_{\mathcal{T}_2}, X^j_{\mathcal{T}_2})  \nonumber \\
    & + \frac{2 \sigma^2}{mn}\sum_{i=1}^m\sum_{j=1}^n k(X^i_{\mathcal{T}_1}, X^j_{\mathcal{T}_2})  \Big)\label{eqn:k_sigma}
\end{align}

Performing classification (or regression) tasks on streaming trees is easily achieved thanks to the explicit expression of the kernel $k_\sigma$ as per equation (\ref{eqn:k_sigma}). In this paper we are interested in binary classification of streaming trees as non-malicious (class $0$) or malicious (class $1$) events. We are given a dataset of input-output pairs $\{\mathcal{T}_i, y_i\}_{i=1}^M$ where the inputs $\mathcal{T}_i$ are streaming trees and the outputs $y_i$ are in $\{0,1\}$. To carry out the classification we make use of a \emph{support vector machine} (SVM) classifier \cite{wang2005support} equipped with the kernel $k_\sigma$ on streaming trees of equation (\ref{eqn:k_sigma}). The binary SVM classification algorithm
aims at solving the following minimisation
\begin{equation}
    \min_{f \in \mathcal{H}_{k_\sigma}} \sum_{i=1}^M L(y_i, f(\mathcal{T}_i)) + \lambda ||f||_{\mathcal{H}_{k_\sigma}}
\end{equation}
where $L(y_i,f(x_i)) = \max(0,1-y_if(x_i))$, and $\lambda$ is the penalty hyperparameter. Following \cite{scholkopf2018learning}, the optimal solution to this minimisation can be expressed in terms of the kernel $k_\sigma$ as follows: for any streaming tree $\mathcal{T}$
\begin{equation}
    f^*(\mathcal{T}) = sgn\Big(\alpha_0 + \sum_{i=1}^My_i \alpha_i k_\sigma(\mathcal{T}, \mathcal{T}_i)\Big)
\end{equation}
where $\alpha_i$ are scalar coefficients computed from solving a quadratic programming problem. Our algorithm, that we call SK-Tree, consists of an SVM classifier that uses the Gram matrix associated to the kernel $k_\sigma$ and computed via Algorithm \ref{algo:algo}. In the latter, $\mathrm{PDESolve}$ stands for a call to a PDE solver to evaluate the signature kernel $k$ (see Sec. \ref{ssec:signaturekernel}) and the notation $0_{m \times n}$ indicates the zero-matrix in $\mathbb{R}^{m \times n}$. We implemented the full SK-Tree algorithm as a ready-to-use estimator using the popular python library scikit-learn \cite{scikit-learn} and we make it publicly available at \url{https://github.com/crispitagorico/SK-Tree}. 

\begin{algorithm}[h]
    \caption{SK-Tree Gram matrix}
    \label{algo:algo}
    \begin{algorithmic}[1]
         \State\textbf{Input: } $M$ str. trees $\left\{\mathcal{T}_i = \left(X_{\mathcal{T}_i}^{1}, ..., X_{\mathcal{T}_i}^{k_i}\right)\right\}_{i=1}^M$, $\sigma>0$
         \State Initialize $G \leftarrow 0_{M \times M}$
            \For{each pair $(i,j) \in \{1,...,M\}^2$} 
            \State{Initialize $K_1 \leftarrow 0_{k_i \times k_i}, K_2 \leftarrow 0_{k_j \times k_j}, K_3 \leftarrow 0_{k_i \times k_j}$}
            \vspace{0.05cm}
            \For{$(p,q) \in \{1,...,k_i\}^2$}
            \State $K_1[p,q] \leftarrow \mathrm{PDESolve}(X_{\mathcal{T}_i}^{p},X_{\mathcal{T}_i}^{q})$
            \EndFor
            \For{$(p,q) \in \{1,...,k_j\}^2$} 
            \State $K_2[p,q] \leftarrow \mathrm{PDESolve}(X_{\mathcal{T}_j}^{p},X_{\mathcal{T}_j}^{q})$
            \EndFor
            \For{$(p,q) \in \{1,...,k_i\} \times \{1,...,k_j\}$}
            \State $K_3[p,q] \leftarrow \mathrm{PDESolve}(X_{\mathcal{T}_i}^{p},X_{\mathcal{T}_j}^{q})$
            \EndFor
            \State $G[i,j] \leftarrow \frac{\mathrm{sum}(K_1)}{k_i(k_i-1)} + \frac{\mathrm{sum}(K_2)}{k_j(k_j-1)} - 2\times \frac{\mathrm{sum}(K_3)}{k_ik_j}$
            \EndFor
            \State $G\leftarrow\exp(-\sigma^2 G)$ 
            \State\textbf{Output:} The Gram matrix $G$. 
    \end{algorithmic}%
\end{algorithm}

\section{Data and experiments}\label{sec:experiments} 

The Operationally Transparent Cyber (OpTC) dataset \cite{optc} was released by DARPA in June 2020.  It consists of data collected from an isolated network of 1000 hosts over a multi-day period.  We use OpTC's \emph{ecar} data, which records endpoint activity in an extended CAR format based on MITRE's CAR data model \cite{mitrecar}.  For example, a process creation event appears as follows (some fields omitted for clarity):

\begin{verbatim}
{"action":"CREATE",
"actorID":"437acfc7-d9ef-4c60-a108-...",
"hostname":"SysClient0201.systemia.com",
"object":"PROCESS",
"objectID":"b9d06a48-0968-4bda-b743-...",
"properties":...,
"timestamp":1569245579591}
\end{verbatim}

Here \texttt{actorID} and \texttt{objectID} uniquely identify the parent and child processes.  In events of other types, such as file and network activity, the \texttt{actorID} similarly provides a unique identifier of the process responsible for the event.

\subsection{Modelling the data as streaming trees}

These events can be interpreted as trees in a natural way, where each branch follows the events that are associated with a particular process.  At a process creation event, the tree splits into two branches: one following the continuing parent process, the other following the new child process. Specifically, we produce a tree in 23 dimensions:
\begin{itemize}
\item The first dimension represents time.
\item The next two dimensions encode the structure of the process tree: they represent the depth in the tree, and the number of processes that this branch has set off.  At a process creation event, the child branch and parent branch immediately increment in these dimensions respectively.
\item We take 20 other types of events defined by (object, action) pairs, eg (MODULE, LOAD) or (FILE, DELETE).  Each event type is associated a dimension.  Whenever an event of this type is observed along a branch, the branch increments in the appropriate dimension.
\end{itemize}



The method easily extends to other sources of host data: the only requirement is that process creation events are logged (including the identity of the parent and child processes), enabling a process tree to be built.  Even in the OpTC \emph{ecar} data there is much more information available in the \texttt{properties} of each event: e.g. the identities of processes and files; source and destination ports of network connections.  We currently ignore these properties, but they could be used to select a set of event types that better summarises the data.

\subsection{Results}

The vast majority of activity in the OpTC dataset is benign: this activity continues throughout the period.  However on three days, red-team attackers are active on the network, providing a source of known malicious activity that we aim to detect.  The attackers' actions are detailed in a PDF document that accompanies the OpTC dataset.  We take all processes that correspond to activity listed in the document, together with their descendants, and mark these processes as malicious.

\begin{figure}
    \centering
    \includegraphics[scale=0.4]{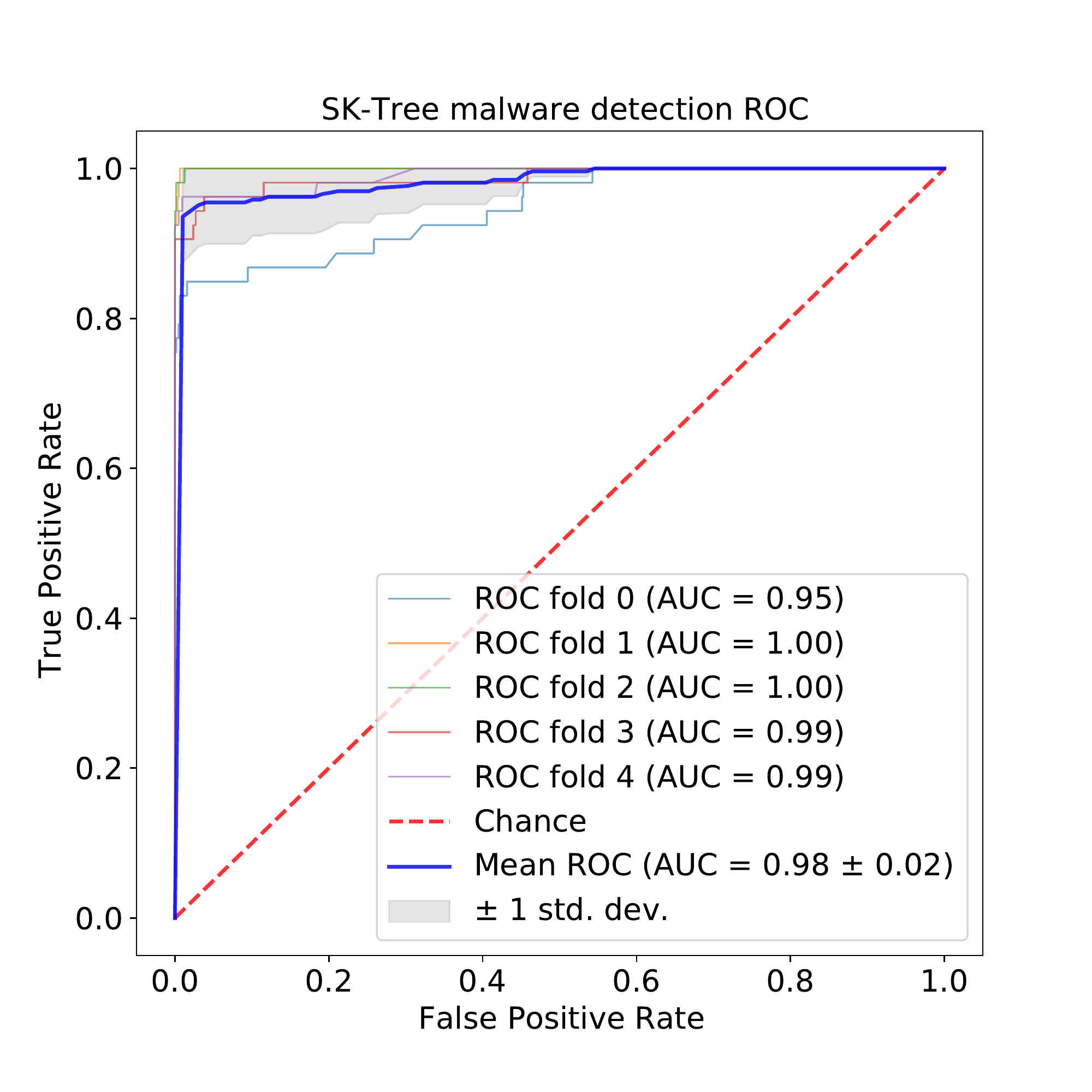}
    \caption{ROC evaluation of the SK-Tree binary classifier on the OpTC data}
    \label{fig:roc}
\end{figure}

To form the trees, we take all activity on the first day of malicious activity from host 0201 (since this host has the greatest proportion of malicious activity).  We split each process into 15-minute windows of activity and create a streaming tree for each.  The associated label is 1 if the root process is malicious, 0 otherwise.  All branches are scaled to have mean 0, sd 1.  We discard any tree with fewer than 2 or more than 200 events. As a result, the final dataset contains $4\,199$ streaming trees. The SK-Tree binary classifier is run on random $5$ fold train-test partitions and we report the mean and standard deviation of the AUROC score on Fig. \ref{fig:roc}. The hyperparameters of SK-Tree are selected by cross-validation via a grid search on the training set of each fold. As it can be observed, we achieve a $98 \%$ AUROC score. 

\section{Conclusion}
In this paper we introduced the notion of a \emph{streaming tree} to describe computer process activity as a generic data structure that encompasses most of the complex, hierarchical, sequential and multimodal nature of the data involved. We then introduced SK-Tree, a new supervised learning method for systematic malware detection on streaming trees that is robust to the irregular sampling and high dimensionality of the underlying streams. SK-Tree is based on the signature kernel, which recently emerged as a leading machine learning tool for learning with complex sequences of data. We finally demonstrated the effectiveness of SK-Tree at detecting malicious events on a portion of the publicly available OpTC dataset \cite{optc} achieving a AUROC score of $98\%$.

\section*{Acknowledgement}

C.S. was supported by the EPSRC grant EP/R513295/1. M.L. was supported by the OxWaSP CDT EPSRC grant EP/L016710/1. All authors were supported by the Alan Turing Institute under the EPSRC
grant EP/N510129/1 and the Defence and Security Programme and by DataSig under the EPSRC grant EP/S026347/1.

\bibliographystyle{IEEEtran}
\bibliography{references}

\end{document}